\documentclass{mn2e}
\input epsf

\newcommand{\Fs}{\,^*\! F}

\newcommand{\bJ}{\bmath{J}}

\newcommand{\bB}{\bmath{B}}
\newcommand{\bE}{\bmath{E}}
\newcommand{\bH}{\bmath{H}}

\newcommand{\bS}{\bmath{S}}

\newcommand{\text}[1]{\quad\mbox{#1}\quad}

\newcommand{\vpr}[2]{\bmath{#1} \!\times\! \bmath{#2}}

\newcommand{\vcurl}[1]{\vpr{\nabla}{#1}}

\newcommand{\pder}[2]{\frac{\partial #1}{\partial #2}}
\newcommand{\Pd}[1]{\partial_{#1}}

\newcommand{\ortt}[1]{ \bmath{i}_{\hat{#1}} }

\newif\ifAMStwofonts

\title
{General relativistic MHD simulations of monopole magnetospheres of
black holes}
\author[S.S. Komissarov]
{
  S.S.Komissarov\\
Department of Applied Mathematics, 
The University of Leeds, 
Leeds LS2 9JT }

\begin{document}
\label{firstpage}
\maketitle

\begin{abstract} 

In this paper we report the results of the first ever 
time-dependent 
general relativistic magnetohydrodynamic simulations of the 
magnetically dominated monopole magnetospheres of black holes. 
It is found that the numerical solution evolves towards a 
stable steady-state  solution which is very close to the 
corresponding force-free solution found by Blandford and Znajek. 
Contrary to the recent claims, the particle inertia does not 
become dynamically important near the event horizon and the 
force-free approximation provides a proper framework for 
magnetically dominated magnetospheres of black holes. 
For the first time, our numerical simulations show the development 
of an ultra-relativistic particle wind from a rotating black hole. 
However, the flow remains Poynting dominated all the way 
up to the fast critical point. This suggests that the details 
of the so-called ``astrophysical load'', where the electromagnetic 
energy is transferred to particles, may have no effect on the 
efficiency of the Blandford-Znajek mechanism.  
\end{abstract}


\section{Introduction}

It is now widely believed that the relativistic jets generated in
active galactic nuclei, galactic microquasars and, presumably, during
gamma ray bursts are powered by rapidly rotating black holes.  This
paradigm is largely based on the theoretical results of Blandford and
Znajek (1977), who argued that, under typical conditions of
astrophysical black holes, their rotational energy can be efficiently
extracted in the form of magnetically dominated relativistic wind.

In fact, the original model of Blandford and Znajek was based on the
approximation of steady-state force-free degenerate electrodynamics
(FFDE, see also Macdonald \& Thorne 1982).  In this approximation the
dynamical role of magnetospheric plasma is reduced to providing
perfect conductivity in the space surrounding the black hole, and the
particle inertia is totally neglected.  Not all mathematical and
physical aspects of this electrodynamic model are well
understood. Similar classical systems, e.g. the Faraday disc, involve
the poloidal current driven over the surface of a rotating conductor
by a non-electrostatic electromotive force. Moreover, the inertia of
this conductor is a key factor in the origin of this force. Obviously,
there is no such conductor in the Blandford-Znajek model. Often, this
missing element of the model is artificially introduced in the form of
the so-called ``membrane'', or the ``stretched horizon'', located
somewhat above the real event horizon \cite{MT82,TPM}. Mathematically,
this is consistent with the so-called ``horizon boundary conditions''
employed in the Blandford-Znajek model \cite{Z77}.

However, the ``Membrane paradigm'' cannot always provide correct
insights into the black hole electrodynamics (this would be a great
mystery if it did.)  Since the membrane is very close to the real
horizon, the outgoing fast waves emitted by the membrane are highly
red-shifted. Moreover, the membrane cannot at all communicate with the
outer space by means of Alfv\'en waves because the inner Alfv\'en
surface is quite far away from the horizon.  As the result, these
waves cannot transport angular momentum and adjust the poloidal
electric current of the outgoing wind as they do so, for example, in
stellar winds of magnetised stars. Punsly and Coroniti (1990) used
these causality arguments to expose the lack of physical clarity in
the electrodynamic model of Blandford and Znajek (1977) and its
representation in the membrane paradigm.  In fact, they concluded that
there is no physically meaningful electromotive force in the
Blandford-Znajek model but only the artificial one which has been
effectively introduced via Znajek's horizon condition.  If so, then
the Blandford-Znajek model and, hence, its steady-state solutions have
to be nonphysical.  Since these FFDE solutions are, in fact, proper
mathematical solutions their nonphysical nature has to give itself away
via instability \cite{Pun-Cor,Pun01}.

In fact, more or less the same criticism was applied to the GRMHD
model of black hole magnetospheres developed by Phinney (1982) as it
also utilized Znajek's horizon condition.  In addition to this
criticism, Punsly and Coroniti developed a rather attractive
alternative GRMHD theory of black hole magnetospheres where particle
inertia played the key role in the extraction of rotational energy of
black holes \cite{Pun-Cor1,Pun01}.  While almost everywhere in their
models the magnetic field was completely dominating, in some regions
within the black hole ergosphere the plasma particles were accelerated
to such a high Lorentz factor that their inertia could no longer be
ignored.  This led to strong inertially driven electric currents
flowing across the magnetic field lines, so that the system resembled
the famous Faraday disc rather closely. 
 
In spite of the great astrophysical importance of the
electrodynamic/MHD mechanism of extraction of rotational energy of
black holes and admirable efforts of B.Punsly, the dramatic clash
between the Blandford-Znajek and Punsly-Coroniti models seemed to
remain largely unnoticed by the astrophysical community for a whole
decade.  Only very recently, following the developments in the theory
of force-free electrodynamics \cite{Uch,Gru,Kom02a} and the impressive
progress in numerical methods for relativistic astrophysics, e.g.
\cite{Pons,Kom99,Koi99,Koi03,Kol,Gam,Del-Zan,Dev-Haw}, things began 
to change.
One of most interesting recent results is concerned with the
conjecture of Punsly and Coroniti on instability of the
Blandford-Znajek monopole solution. In \cite{Kom01b} this conjecture
was tested by means of time-dependent general relativistic FFDE
simulations and was found to be incorrect. Contrary to the appealing
Punsly-Coroniti causality arguments, this steady-state monopole
solution of Blandford and Znajek is asymptotically stable.  There is,
however, the question of validity of the FFDE approximation which has
to be fully investigated before reaching any final conclusion.  

There are many examples in physics and astrophysics where
particular approximations fail to describe certain natural phenomena
because one or another physical factor ignored in the approximation
becomes important. These examples taught us always to check the limits
of applicability of our theoretical models and never be satisfied by
their mere self-consistency. The very relevant example can be found in
the theory of magnetically dominated pulsar winds where particle
inertia becomes an important factor as a result of acceleration by
electromagnetic field (Mestel 1999, Section 13.2.3) Other potential
examples in the theory of black hole magnetospheres are discussed by
Punsly (2001, Sections 8 and 9).  Recently, Punsly (2004) argued that
FFDE is deficient near the event horizon where ``the plasma attains an
infinite inertia in a global sense''. Macdonald and Thorne
\shortcite{MT82} also anticipated a breakdown of the FFDE
approximation near the event horizon, though no detailed explanations
were given.  Thus, in the problem of electrodynamic/MHD mechanism of
extraction of rotational energy of black holes we have to verify that
the particle inertia can indeed be ignored everywhere without
oversimplifying this problem.  This can be done via GRMHD simulations
of black hole magnetospheres.

The main goal of this paper is to explore the role of particle inertia
in the pair-filled monopole magnetospheres of black holes by means of
such simulations. In particular, we need to know whether incorporation
of initially small particle inertia will eventually lead to strong
deviations from the corresponding electrodynamic solution both
locally, e.g. near the event horizon, and globally. By latter we mean
changes of the global system of electric current, the angular velocity
of magnetic field lines, and, hence, the efficiency of the energy
extraction.  Since, at present only codes for the perfect relativistic
MHD are readily available, the monopole magnetic configuration appears
to be the most suitable one for this purpose. Indeed, this
configuration allows a dissipation free FFDE solution
\cite{BZ,Kom01b}.  Dipolar magnetospheres, like the one considered by
Koide \shortcite{Koi03}, are less suitable because the corresponding
electrodynamic solution involves a strong current sheet located within
the ergosphere \cite{Kom02b,Kom04}.  This indicates that the
approximation of perfect MHD is also likely to break down in this
configuration.

\section{Basic equations and numerical method}

In this study we use the Kerr-Schild coordinates, which are more
suitable for our purpose than the more popular Boyer-Lindquist
coordinates as they do not introduce a coordinate singularity at the
event horizon.  Not only this allows to increase efficiency of
computer simulations but we may also place the inner boundary of our
computational domain inside the event horizon. At such a boundary we
may confidently impose usual radiative boundary conditions. Thus, the
disputed Znajek's horizon conditions are not involved in any form.  

Using the standard
notation of the 3+1 approach, the metric form of the Kerr-Schild
coordinates, ${t,\phi,r,\theta}$, is

\begin{equation}
  ds^2 = (\beta^2-\alpha^2) dpt^2 + 2 \beta_i dx^i dt +
         \gamma_{ij}dx^i dx^j,
\label{metric}
\end{equation}
where $\gamma_{ij}$ is the metric tensor of space,
\begin{equation}
  \gamma_{ij} = \left(\begin{array}{ccc} \Sigma \sin^2\theta/\kappa^2
       & -a\sin^2\theta(1+Z) & 0 \\ -a\sin^2\theta(1+Z) & 1+Z & 0 \\ 0
       & 0 & \kappa^2 \end{array} \right),
\label{gamma}
\end{equation}
$\alpha$ is the lapse function,
\begin{equation}
    \alpha = 1/\sqrt{1+Z},
\end{equation}
and $\bbeta$ is the shift vector,
\begin{equation}
    \beta^i = (0,\frac{Z}{1+Z},0).
\end{equation}
In these equations
\begin{eqnarray}
  \nonumber \kappa^2 &=& r^2 +a^2\cos^2\!\theta, \\ \nonumber Z &=&
  2r/\kappa^2,\\ \nonumber \Sigma &=&
  (r^2+a^2)^2-a^2\Delta\sin^2\!\theta, \\ \nonumber \Delta &=&
  r^2+a^2-2r.
\end{eqnarray} 
and we assume that indexes 1,2,3 correspond to the $\phi$-, $r$-, and
$\theta$-coordinates respectively.  Notice that we utilize such units
that $G=M=c=1$ and $4\pi$ does not appear in the Maxwell equations.

The system of perfect general relativistic magnetohydrodynamics
(GRMHD) includes the continuity equation,
\begin{equation} 
\Pd{t}(\sqrt{\gamma}\rho u^t)+ \Pd{i}(\sqrt{\gamma}\rho u^i)=0,
\label{cont1} 
\end{equation}
the energy-momentum equations,

\begin{equation} 
\Pd{t}(\sqrt{\gamma}T^t_{\ \nu})+ \Pd{i}(\sqrt{\gamma}T^i_{\ \nu})=
\frac{1}{2} \Pd{\nu}(g_{\alpha\beta}) T^{\alpha\beta} \sqrt{\gamma},
\label{en-mom1} 
\end{equation}
and the induction equation,

\begin{equation} 
\Pd{t}(B^i)+e^{ijk}\Pd{j}(E_k) =0.
\label{ind1} 
\end{equation}
Here $g_{\alpha\beta}$ is the metric tensor of spacetime,
$\gamma=\det(\gamma_{ij})$, \\ $e^{ijk}$ is the Levi-Civita
pseudo-tensor of space, $\rho$ is the proper mass density of plasma,
$u^\nu$ is its four-velocity vector.  The total stress-energy-momentum
tensor, $T^{\mu\nu}$, is a sum of the stress-energy momentum tensor of
matter,
\begin{equation}
   T_{(m)}^{\mu\nu} = wu^\mu u^\nu -p g^{\mu\nu},
\end{equation}
where $p$ is the thermodynamic pressure and $w$ is the enthalpy per
unit volume, and the stress-energy momentum tensor of electromagnetic
field,
\begin{equation}
   T_{(e)}^{\mu\nu} = F^{\mu\gamma} F^\nu_{\ \gamma} -
   \frac{1}{4}(F^{\alpha\beta}F_{\alpha\beta})g^{\mu\nu},
\end{equation}
where $F^{\nu\mu}$ is the Maxwell tensor of the electromagnetic field.
The electric field, $\bE$, and the magnetic field, $\bB$, are defined
via
\begin{equation}
  E_i=\frac{\alpha}{2} e_{ijk}\Fs^{jk} ,
\end{equation}
and
\begin{equation}
  B^i=\alpha \Fs^{it},
\end{equation}
where $\Fs^{\mu\nu}$ is the Faraday tensor of the electromagnetic
field, which is simply dual to the Maxwell tensor. In the limit of
ideal MHD

\begin{equation}
  \bE=-\vpr{v}{B} \text{or} E_i=e_{ijk}v^jB^k,
\end{equation}
where $v^i=u^i/u^t$ is the usual 3-velocity of plasma. All the
components of vectors and tensors appearing in equations
(\ref{cont1},\ref{en-mom1},\ref{ind1}) are the components in the basis
of global coordinates.

Our numerical scheme is a generalization of the scheme 
for the special relativistic MHD described in Komissarov(1999), 
where the magnetic field is evolved using the method of 
constraint transport \cite{Eva-Haw}. It shares many common features 
with other existing numerical schemes for GRMHD and for 
this reason we only briefly outline its design. 

This is a 2-dimensional scheme with enforced axial symmetry.
The set of primitive variables include $P$, $\rho$, 
the components of magnetic field, $B^{\hat{i}}$, and the 
components of fluid four-velocity, $u^{\hat{i}}$,  
as measured in the orthonormal basis of the local fiducial 
observer (FIDO),   
$\{\ortt{t},\ortt{\phi},\ortt{r},\ortt{\theta}\}$. The 
transformation from the coordinate basis to the orthonormal
basis of FIDO is determined by 

\begin{eqnarray} 
\nonumber
\ortt{t} &=& \frac{1}{\alpha} \pder{}{t} -
\frac{\beta^r}{\alpha} \pder{}{r},\\ 
\nonumber
\ortt{\phi} &=& \frac{1}{\sqrt{\gamma_{\phi\phi}}}\pder{}{\phi},\\ 
\ortt{r} &=& {\cal A} \left( \pder{}{r}-
         \frac{\gamma_{r\phi}}{\gamma_{\phi\phi}} \pder{}{\phi} 
           \right),  \\
\nonumber
\ortt{\theta} &=& \frac{1}{\sqrt{\gamma_{\theta\theta}}}
                  \pder{}{\theta} ,
\end{eqnarray}
where 
$
  {\cal A} = \sqrt{\gamma_{\phi\phi}\gamma_{\theta\theta}/
             \gamma} .
$
Notice, that $\rho$, $P$, $u^{\hat{i}}$ and $B^{\hat{\phi}}$ are
defined at the geometrical centers of the computational cells, whereas
$B^{\hat{r}}$ and $B^{\hat{\theta}}$ are defined at the geometrical
centers of the cell interfaces.

At the beginning of each time step both $B^{\hat{r}}$ and
$B^{\hat{\theta}}$ are found in the cell centers via linear
interpolation. The cell centered values are then used to set the
Riemann problems at the cell interfaces using a slope-limited
interpolation.  These problems are then solved using the special
relativistic Riemann solver described in \cite{Kom99}, the solution
being interpreted as the one observed by the FIDO initially located at
the interface.  In order to find the resolved state at the interface,
the relative motion of the interface and its FIDO has to be taken into
account \cite{Pons}. This resolved state allows to compute the
interface fluxes of all conserved quantities , first in the FIDO basis
and then in the coordinate basis via the corresponding transformation
laws. These fluxes are then used to evolved the volume averaged mass,
energy, and momentum densities, as well as $B^\phi$.  The values of
$E_\phi$ in the resolved states at the cell interfaces allow to find
the mean values of $E_\phi$ at the cell edges. These are used to
evolve the magnetic flux through the cell interfaces and, hence, to
update the $r$- and $\theta$-components of the interface magnetic
field, first in the coordinate basis and then in the FIDO's basis via
the corresponding transformation law. Next, the updated $r$- and
$\theta-$components of magnetic field at the cell centers are found
via linear interpolation. Finally, the updated conserved variable are
transformed into FIDO's basis and the updated primitive variables are
found using the same iterative procedure as in the special
relativistic scheme. The second order accuracy in time is achieved via
a half time step as described in Komissarov (1999).

\begin{figure*}
\leavevmode \epsffile[0 0 500 250]{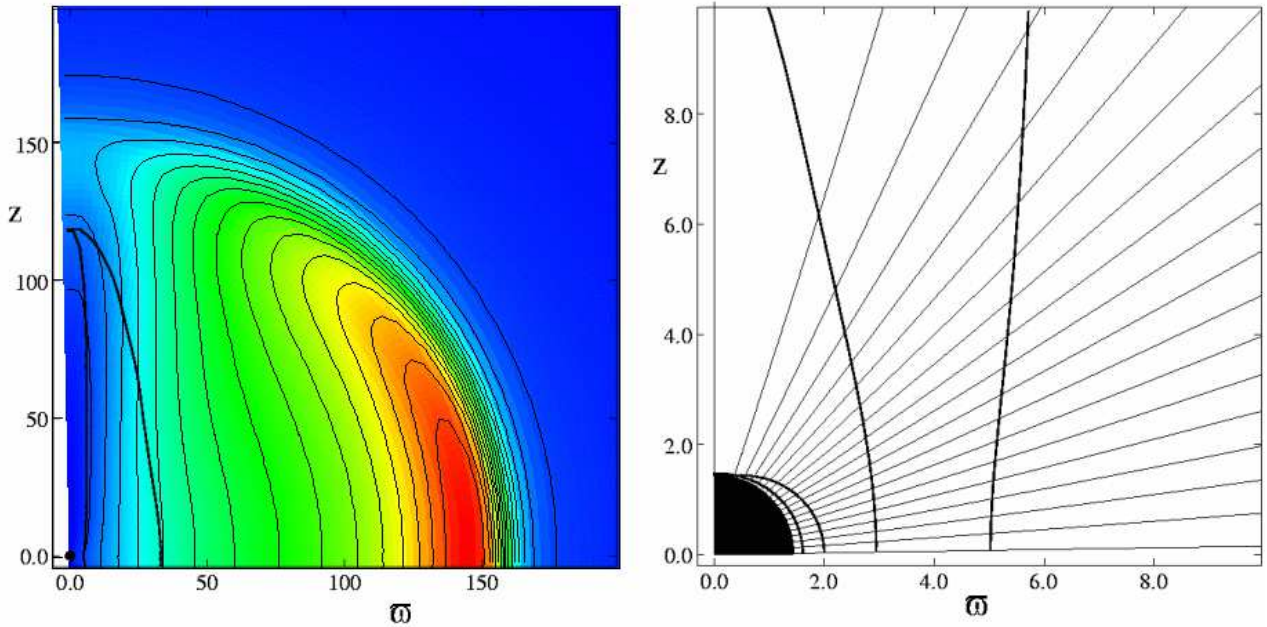}
\caption{Solution for a black hole with $a=0.9$ at $t=170$.  {\it Left
panel:} Thin contour lines show the distribution of the Lorentz
factor, $W$. There are 15 contours equally spaced between $W=1$ and
$W=13.6$. Along the equatorial plane $W$ is gradually increasing until
it reaches a maximum at the cylindrical radius $\varpi\approx 150$.
The thick lines show the Alfv\'en and fast magnetosonic critical
surfaces of the outgoing wind. {\it Right panel:} The inner region of
this solution. The thin lines show the magnetic flux surfaces. The
thick lines show, in the order of increasing distance from the black
hole, the Alfv\'en surface of the ingoing wind, the ergosphere, the
wind separation surface, and the Alfv\'en critical surface of the
outgoing wind.  The inner fast surface is too close to the event
horizon to be seen in this figure.}
\label{sol09}
\end{figure*}

The scheme was tested using the suit described in \cite{Koi99}. In all
cases, where the test problem was based on an existing analytical
solution, the agreement between numerical and analytical solutions 
is very good. However, the suite also includes the ``sub-Keplerian disc'' 
problem for which there is no analytical solution. Surprisingly, our 
numerical solution for this problem is dramatically different
from the one of Koide et al.(1999) -- instead of ``bouncing of
the centrifugal barrier'' our disc gets swallowed by the
hole.  Since the specific angular momentum of this disc is lower 
than the one of the last stable orbit, the ``bouncing'' seems 
highly unlikely and we suspect that the results by Koide et al.(1999) 
are incorrect (in fact, this has been confirmed in private communication 
to the author by K.Shibata.)  In addition, we considered the problem of 
an equilibrium torus around a rotating black hole, both with and without 
an azimuthal magnetic
field. The pure gasdynamic solution is described in \cite{Abr}. The
solution for a magnetized torus does not seem to have been described
in the literature (a paper is being prepared for publication elsewhere.)

\section{Numerical simulations}

In these simulations, the computational grid has 100 cells in the
$\theta$-direction, where it is uniform, and 150 cells in the
$r$-direction. The cell size in the $r$-direction is such that the
corresponding physical lengths in both directions are equal.  The
radiative outer boundary is located far away from the event horizon,
$r_{out} \approx 230$, which ensures that it does not effect the
solution near the black hole.

The initial solution describes a purely radial magnetic field,
\begin{equation} 
\bB = \frac{B_0\sin\theta}{\sqrt{\gamma}} \pder{}{r},
\end{equation}
with $B_0=1$ (this implies a non-vanishing azimuthal component in the
Boyer-Lindquist basis.) The plasma velocity relative to FIDO is set to
zero, $u^{\hat{i}}=0$, whereas its pressure and density are set to the
same value, of $p=\rho=B^2/100$. In any conservative scheme for the
relativistic magnetohydrodynamics, there is an upper limit on the
magnetization of plasma above which the hydrodynamical part of the
solution suffers from large numerical errors \cite{Kom01a,Gam}. At
this limit, which depends on the resolution, the numerical error for
the energy density of the electromagnetic field becomes comparable
with the energy density of matter.  For this reason we could not set
the initial density and pressure to the much lower values typical for
black hole magnetospheres.  Moreover, soon after the start of the
simulations, the numerical solution can be described as a pair of
particle winds, one ingoing and one outgoing (see figure \ref{vel}),
and, as the result, the plasma density gradually slides down towards
the danger zone near, as well as inside, of the wind separation
interface.  To keep the magnetization reasonably low, we had to
continuously pump new plasma in this region. The critical condition we
set in these simulations was
\begin{equation} 
   wW^2 - p = 0.03 B^2,
\label{cond} 
\end{equation}
where $W$ is the Lorentz factor of the flow as measured by FIDO. When
the energy density of matter dropped below $0.03 B^2$, both $\rho$ and
$p$ were artificially increased by the same factor.  To minimize the
effect of the mass injection on the winds kinematics the velocity of
the injected matter was set to be equal to the local velocity of the
wind.  In fact, new particles must be constantly created in real
magnetospheres of black holes but the details of this process can be
rather different \cite{Bes,Hir-Oka,Phi}.

\begin{figure*}
\leavevmode \epsffile[0 0 500 250]{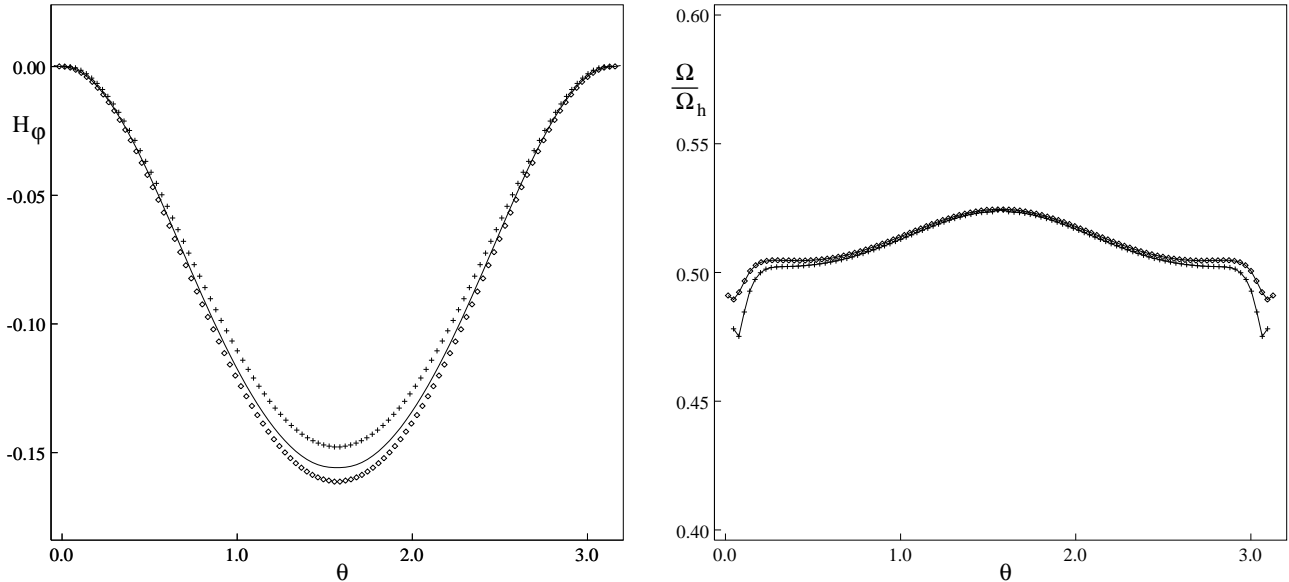}
\caption{ The angular distribution of $H_\phi$ and $\Omega$ for a
black hole with $a=0.9$ at $t=170$. {\it Left panel:} $H_\phi$; the
diamonds show the MHD solution at $r=5$ and the crosses show this
solution at $r=50$. The continuous line show the corresponding FFDE
solution at $r=5$.  {\it Right panel:} $\Omega$; the diamonds show the
MHD solution at $r=5$ and the crosses show this solution at $r=50$.  }
\label{h1}
\end{figure*}

An additional lower limit was set on the value of the thermodynamic
pressure, which was not allowed to drop below $0.01\rho$. In these
simulations we used the polytropic equation of state with the ratio of
specific heats, $\Gamma=4/3$.

Figure \ref{sol09} shows the numerical solution for a black hole with
$a=0.9$ at $t=170$. In the left panel of this figure, where the
distribution of the Lorentz factor as measured by the local FIDO is
shown, one can see an almost spherical wave front which designates the
expanding boundary of the outgoing wind.  This wind is ultra-relativistic 
and superfast within most of its volume. Moreover, the positions of both the fast
and the Alfv\'en critical surfaces in the equatorial plane do no
longer show any noticeable variation at this point.  The wind is
magnetically dominated but its magnetization is slowly decreasing with
distance.

While the Lorentz factor of the outgoing wind gradually increases with
the distance from the black hole, and so does the inertia of the
accelerated particles, the Lorentz factor of the ingoing wind remains
lower than $W=2$ all the way down to the inner boundary of the
computational domain (see also figure \ref{vel}). Thus, the
electromagnetic field does not push the plasma of the ingoing wind
onto the high-velocity orbits and its inertia does not become a key
factor in the flow dynamics. On the contrary, new particle have to be
constantly injected into this wind to keep its inertia above the low
level given by eq.\ref{cond}.  The region of particle injection is
elongated along the symmetry axis with the major semi-axis about 20
and the minor semi-axis about 6.
      
The right panel of this figure shows the key surfaces in the inner
part of this solution where it has settled to a steady-state. The
comparison with fig.1b in \cite{Kom01b} reveals that the locations of
the Alfv\'en surfaces are very close to those in the corresponding
FFDE solution.  As expected, the surface separating the ingoing wind from the
outgoing one is located between the Alfv\'en surfaces \cite{Tak}.

Figure \ref{h1} shows the distribution of the angular velocity of
magnetic field lines, $\Omega$, and the $H_\phi$-component of vector
$\bH$ defined via
\begin{equation}
   H_i=\frac{\alpha}{2}e_{ijk}F^{jk}.
\end{equation}
In steady-state
\begin{equation}
   \vcurl{H}=\bJ,
\end{equation}
where $\bJ$ is the electric current density \cite{Kom04} and, thus,
$H_\phi$ is a measure of the poloidal current.  (In Blandford \&
Znajek (1977) $H_\phi$ is denoted as $B_T$.)  $\Omega$ and $H_\phi$
are very important quantities as they determine the poloidal fluxes of
the electromagnetic energy and angular momentum \cite{BZ,Kom04}.  For
example, the energy flux density, $\bS_p$, is given by
                                                                                
\begin{equation}
   \bS_p = -(H_\phi \Omega) \bB_p,
\label{a14}
\end{equation}
where $\bB_p$ is the poloidal component of the magnetic field.  As one
can see in figure \ref{h1} the distributions of $\Omega$ and $H_\phi$
in our magnetically dominated MHD solution are very close to those
found in the FFDE solution \cite{Kom01b}, all the way up to the fast
critical surface of the outgoing wind. Thus, the particle inertia does
not have much of an effect on the process of extraction of energy and
angular momentum, which remains essentially electromagnetic.

Figure \ref{vel} shows the velocity distribution in the equatorial
plane near the black hole. At the point separating the inflow from the
outflow the angular velocity of plasma, $v^\phi =d\phi/dt$, equals to
the one of the magnetic field lines.  Closer to the black hole
$v^\phi>\Omega$ and near the event horizon it even exceeds the angular
velocity of the black hole, $\Omega_h$.  (Notice, that it is only the 
Boyer-Lindquist $v^\phi$ which always equals to $\Omega_h$ 
at the event horizon.)  
 
Figure \ref{vel} also shows that the Lorentz factor of the ingoing
wind remains very low and, thus, there is no any reason for the
breakdown of the FFDE approximation at the event horizon suggested in
\cite{Pun04,MT82}.
 
\begin{figure}
\leavevmode \epsffile[0 0 250 250]{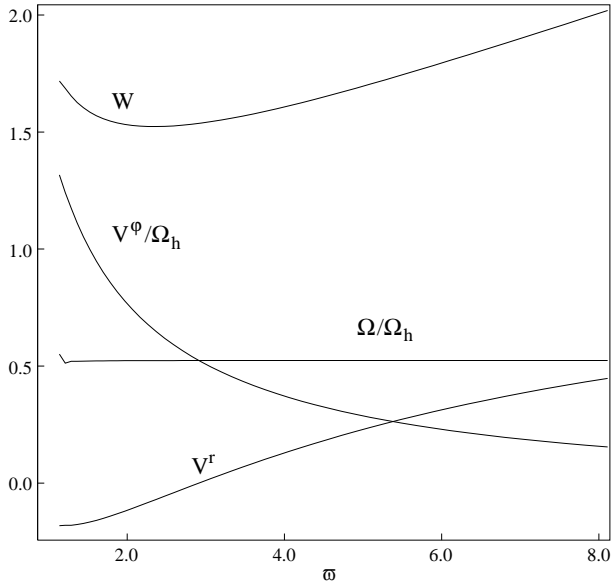}
\caption{The Lorentz factor as measured by FIDO, $W$, the radial
velocity of plasma, $v^r=dr/dt$, the angular velocity of plasma,
$v^\phi=d\phi/dt$, and the angular velocity of magnetic field lines,
$\Omega$, in the equatorial plane at $t=170$.  }
\label{vel}
\end{figure}

\section{Discussion and Conclusions}

In this paper we described the first ever GRMHD simulations of 
monopole  
magnetospheres of rotating black holes. Given the central role 
played by the monopole problem in the development of the general 
theory of black hole magnetospheres it is not surprising that the 
results of these simulations allow to re-examine a number of 
important issues of the theory.  
    
The main conclusion which follows from these results is the validity 
of the FFDE approximation at least in the case of the monopole 
magnetic configuration considered by  Blandford and Znajek(1977).  
Inertia of initially rarefied magnetospheric plasma does 
not grow dynamically important neither in any localized  regions near 
the black hole horizon nor in a global sense. The system of poloidal 
electric current and the efficiency of energy extraction are basically 
the same as in our GRMHD solution as in the FFDE solution found 
earlier \cite{Kom01b}.  This means that in the magnetospheres of black 
holes there exists an electromotive force that has nothing to do with 
particle inertia.  In a separate paper \cite{Kom04} we show         
that, in great contrast to the Faraday disc or a magnetized stellar
wind, the poloidal currents in the Blandford-Znajek model are driven
by the so-called ``gravitationally induced'' electric field, which was
first discovered by Wald (1974). Contrary to the conclusion reached 
in \cite{Pun-Cor}, this field cannot be screened within the black 
hole ergosphere by any static distribution of electric charge.    
We also show there that Znajek's horizon condition is not a boundary 
condition after all, but a regularity condition imposed at the fast 
critical point of the ingoing wind in the limit of vanishing 
particle inertia. This proves the legitimacy of its utilization in 
the steady-state solutions by Blandford-Znajek(1977) and 
Phinney(1982).  

In addition, the numerical solution shows a number of other interesting 
features that deserve discussing. 

Our GRMHD solution remains very close to the FFDE one all the way up 
to the fast critical surface of the outgoing wind and even at the fast 
critical surface the wind is still Poynting flux dominated. Further 
away the electromagnetic energy may be transferred to the wind  
particles. However, the details of this energy transfer as well as the 
details of the interaction between the wind and its surrounding which  
is responsible for the observational phenomena like superluminal jets, 
radio galaxies etc. cannot effect the wind solution in the sub-fast 
region. This makes us wonder whether the key global properties of the 
Blandford-Znajek mechanism, such as its efficiency, are sensitive at all 
to the nature of the so-called ``astrophysical load'' \cite{MT82,TPM}. 

Given the potential importance of this finding let us show that 
a RMHD flow may indeed stay Poynting flux dominated at the fast critical 
point. For this purpose we consider a one dimensional flow in flat 
spacetime. In the limit of cold MHD, the ratio kinetic energy flux 
to the Poynting flux is 
$$
  \kappa = \rho W^2/B_t^2, 
$$  
where $B_t$ is the tangential component of magnetic field. At the 
fast point the flow velocity 
$$
  v^2=b^2/(b^2+\rho), 
$$
where $b^2 = B_n^2 + (B_t/W)^2$, $B_n$ being the normal component 
of the magnetic field. From these two equations one finds that at the 
fast point
\begin{equation}
  \kappa = \frac{1}{W^2} + \frac{B_n^2}{B_t^2}.  
\end{equation}
Thus, the flow is Poynting dominated provided the fast speed 
is ultra-relativistic and the magnetic field is predominantly 
tangential.  

These are the first GRMHD simulations where an ultrarelativistic 
outflow is produced by a black hole. In all previous simulations 
of this sort
\cite{Koi99,Koi00,Koi03,Kom01a,Gam,Dev-Haw} 
no such high velocity outflows were reported,  which 
was a bit worrying given the well known results of various 
astrophysical observations. In more realistic simulations of the 
future, which will include both the accretion disc and its 
magnetised corona,  it may be possible to obtain not only 
ultra-relativistic but also well collimated outflows.    

Since the fast and the Alfven surfaces of the outgoing wind are 
relatively far away from the event horizon, one 
may expect the outgoing wind to be more or less accurately described 
by the flat space monopole solution. According to \cite{Bes97} 
the ratio of the fast suface radius, $r_f$, and the Alfv\'en
surface radius, $r_a$, in the equatorial plane of this solution is   
$$
  r_f/r_a \approx \sigma^{1/3},
$$
whereas the Lorentz factor at the fast surface 
$$
   W_f=\sigma^{1/3}, 
$$ 
where $\sigma$ is the magnetization papameter. In our simulations 
$\sigma^{1/3} \approx 4.08$, $r_f/r_a=6.60$, and $W_f=4.35$. Thus, 
the numerical and the analytical results agree quite well.      

The presented results also suggest a way of dealing
with such more realistic problems of computational astrophysics 
of black holes which involve both magnetically dominated  
ultrarelativistic components and  particle dominated slow 
components.  Namely, one may use the same GRMHD approximation to
model all components of the system but impose an upper limit on 
the magnetization of the ultrarelativistic component, via appropriate 
floor values for both the pressure and density of matter. Although, 
there is nothing particularly new in such an approach and utilization 
of similar lower/upper limits is widely spread in computational 
practice, the accurate representation of magnetically dominated 
component in our simulations is reassuring.

\end{document}